\documentclass[12pt]{article}
\usepackage{graphicx}
\usepackage{setspace}
\usepackage{epstopdf}
\usepackage{bm}
\usepackage{float}
\usepackage{amssymb}
\usepackage{amsmath}

\newcommand{\ket}[1]{ | #1 \rangle}

\newcommand{\ua}{\uparrow}
\newcommand{\da}{\downarrow}

\begin{document}

\title{New frontiers with quantum gases of polar molecules}
\author{Steven A. Moses$^1$, Jacob P. Covey, Matthew T. Miecnikowski, Deborah S. Jin, Jun Ye$^{\dagger}$\\
\\
\normalsize{JILA, National Institute of Standards and Technology and University of Colorado,} \\ \normalsize{and Department of Physics, University of Colorado, Boulder, CO 80309, USA}
\\
\normalsize{$^1$Present address: Joint Quantum Institute, University of Maryland Department of Physics and National} \\
\normalsize{Institute of Standards and Technology, College Park, MD 20742, USA} \\
\vspace{-0.8cm}
\normalsize{Corresponding author: $^{\dagger}$ye@jila.colorado.edu}}
\date{}
\maketitle

\textbf{The field of ultracold quantum matter has burgeoned over the last few decades, thanks to the growing capabilities for atomic systems to be probed and manipulated with exquisite control.  Researchers can now precisely create and study quantum many-body states that are effectively isolated from the external environment.  Much of the work in ultracold matter has focused on systems of alkali or alkaline-earth atoms, mainly due to their ease of cooling. Extending this precise control to molecules has seen rapidly increasing interest and activity, as molecules possess additional degrees of freedom that make them useful for tests of fundamental physics, studying ultracold chemistry and collisions, and engineering qualitatively new types of quantum phases and quantum many-body systems.  Here, we review one particularly fruitful research direction: the creation and manipulation of ultracold bialkali molecules. The recent success in creating a quantum gas of polar molecules opens many exciting research opportunities.}
\\

Atomic, Molecular, and Optical (AMO) systems offer experimentalists the ability to precisely control interactions in a quantum many-body system \cite{Bloch2012, Chin2010}, and a rich set of experiments has already been performed.  Some prominent examples include studies of the BEC-BCS crossover in two-component Fermi gases \cite{Regal2004, Zwierlein2004, Bartenstein2004} and the superfluid-Mott insulator transition in ultracold bosons loaded into a 3D optical lattice \cite{Greiner2002}.   Neutral atomic systems normally have point-like contact interactions that can be parametrized with a single quantity, the scattering length $a$, which can be tuned via a Feshbach resonance \cite{Chin2010}.  In recent years, systems with long-range and anisotropic interactions have garnered much attention. One natural example is the dipolar interaction between polar molecules, which is the focus of this Review. Of course, many other experimental platforms are also being pursued that feature long-range interactions, including highly magnetic atoms \cite{Lu2010, Aikawa2012}, trapped ions \cite{Blatt2012}, superconducting circuits \cite{Devoret2013}, Rydberg atoms \cite{Low2012}, and atoms coupled to photonic structures \cite{Douglas2015}.

The dipolar interaction exhibits several important attributes. First, the energy scales in dipolar systems are usually much larger than those in typical atomic alkali systems, making it easier to study interaction-driven structural properties and non-equilibrium dynamics. Second, and perhaps more importantly, the anisotropic and long-range nature of the interaction allows for the realization of novel quantum phases, especially when the spatial dimensions of the system can be varied with optical lattices. Some of these engineered systems may find relevance to outstanding problems in condensed-matter physics \cite{Baranov2012, Yao2014}; moreover, qualitatively new types of physics can emerge in the presence of long-range interactions \cite{Baranov2002, Cooper2009, Pikovski2010, Kuns2011, Knap2012, Yao2013, Syzranov2014}.  As a specific example of a unique feature of long-range interactions, a spin-1/2 Hamiltonian can be encoded based on a pair of opposite-parity rotational states where dipolar interactions give rise to a direct spin exchange coupling between molecules \cite{Gorshkov2011, Hazzard2013}. With this scheme implemented in an optical lattice, many-body spin dynamics can be observed where the spin and motional degrees of freedom are completely decoupled, which greatly reduces the entropy requirements compared to neutral atoms for realizing quantum magnetic models.  Ultracold molecules can also be used for many other applications, including ultracold chemistry and precision measurements, and a general overview of current research directions with polar molecules is shown in Fig.~1.

Production of a quantum degenerate gas of molecules requires ultralow temperatures and high phase-space densities.  For alkali atoms, one can take advantage of nearly-closed optical cycling transitions that allow for efficient laser cooling to $1-100 \,  \mu$K temperatures \cite{Phillips1998}, which can be followed with evaporative cooling to bring the gas to quantum degeneracy \cite{Cornell2002, Ketterle2002}.  Molecules have much richer internal structure, with many rovibrational states associated with each electronic potential.  This complexity makes it difficult to directly cool molecules or to accumulate a significant fraction of molecules in a single quantum state.  For example, the current state-of-the-art for the molecular magneto-optical trap (MOT) \cite{Stuhl2008, Hummon2013, Barry2014} is about 2000 molecules of SrF at $\sim 400 \, \mu$K \cite{Norrgard2016}, corresponding to a phase-space density orders of magnitude smaller than typical atomic alkali MOTs of $10^9-10^{10}$ atoms at $\sim 10 \, \mu$K.  Many other techniques have also been employed to create cold molecules, such as photoassociation \cite{Jones2006}, buffer gas cooling \cite{Hutzler2012}, Stark deceleration \cite{Meerakker2008, Stuhl2012}, and Sisyphus cooling \cite{Prehn2016}; however, the achieved phase-space densities are all very far from that required for quantum degeneracy.  In this Review we discuss the only method demonstrated so far for producing ultracold molecules in the quantum regime: the creation of weakly bound bialkali dimers from an ultracold atomic mixture, followed by coherent optical state transfer to the rovibrational ground state.  Since there are only two stable fermonic alkali atoms, $^6$Li and $^{40}$K, any fermionic molecule must have one of these atoms, along with a boson.  Molecules comprised of two fermions or two bosons are bosonic.  This approach has been extremely successful in creating ultracold, dense samples of ground-state molecules, including fermionic KRb \cite{Ni2008}, bosonic Cs$_2$ \cite{Danzl2010}, bosonic RbCs \cite{Takekoshi2014, Molony2014}, fermionic NaK \cite{Park2015}, and bosonic NaRb \cite{Guo2016}.  Our recent work finally demonstrated the production of a quantum degenerate molecular gas in an optical lattice \cite{Moses2015}.

\textbf{Creating ultracold molecules from ultracold atoms.} In the first experimental successes reported for heteronuclear molecular KRb~\cite{Ni2008} and homonuclear molecular Cs$_2$~\cite{Danzl2010}, researchers started with quantum degenerate atomic gases and created ultracold molecules in the absolute rovibrational ground state of the electronic ground (singlet) potential via a coherent process that consists of two main steps.  First, a Feshbach resonance is used to convert pairs of free atoms to weakly bound molecules \cite{Donley2002, Jochim2003, Herbig2003, Kohler2006}.  Then, these weakly bound molecules are transferred coherently to the rovibrational ground state via STImulated Raman Adiabatic Passage (STIRAP) \cite{Bergmann1998}.  Homonuclear Rb$_2$ molecules in the ground state of a triplet electronic potential were created using the same protocol \cite{Lang2008}.  In certain molecules, such as Sr$_2$, very favorable Franck-Condon factors allow for efficient photoassociation into a deeply-bound state \cite{Zelevinsky2008, Reinaudi2012} or for direct STIRAP from two atoms on individual sites of a 3D lattice \cite{Stellmer2012}.

In the simplest case, STIRAP involves two lasers: one couples the Feshbach molecule state $\ket{f}$ to an excited state $\ket{e}$ (which is usually lossy), and another couples $\ket{e}$ to a deeply bound state $\ket{g}$.  The scheme may be more complicated, as in Cs$_2$, where there are actually five states involved, and either two stages of STIRAP or a four-photon scheme are required \cite{Danzl2010}.  The state $\ket{e}$ is chosen to optimize the Franck-Condon factors with both $\ket{f}$ and $\ket{g}$. The Franck-Condon overlap between a weakly bound Feshbach molecular state and selected excited molecular states is much higher than for free atoms.  For the two-photon transition to have a reasonable strength, the excited molecular state must also have an admixture of singlet and triplet characters (since $\ket{g}$ is a singlet while $\ket{f}$ is a triplet).  Bialkalis typically have spin-orbit coupling in excited electronic states that gives rise to singlet-triplet mixing.  Fig.~2a displays the relevant energy levels for KRb.  The overall structure is very similar for other bialkali molecules.

STIRAP transfers the entire population from $\ket{f}$ to $\ket{g}$ without populating the lossy state $\ket{e}$, provided that the two laser fields are phase coherent throughout the transfer process.  The time dependence of the laser intensities (Fig.~2b) is chosen to maintain a coherent dark state that adiabatically evolves from the initial to the final state.  Photons carry away the binding energy, and the coherent process does not produce any heating.  For typical experimental parameters, a relative laser linewidth less than 1 kHz is required for efficient transfer, which can readily be achieved by stabilizing lasers to an optical frequency comb \cite{Ni2008, Ospelkaus2008} or a high-finesse optical cavity \cite{aikawa2009, gregory2015}.  Magnetic field noise must also be minimized, since Feshbach molecules usually have an appreciable magnetic dipole moment (NaRb is a notable exception \cite{Guo2016}), while singlet ground-state molecules have a negligible magnetic dipole moment.

\textbf{Manipulating rotational and hyperfine states.} The ground state of bialkalis is an electronic singlet, but hyperfine structure is inherited from the nuclear spins of the constituent atoms.  Angular momentum selection rules allow STIRAP to be configured to populate only one of these hyperfine states, thus creating a spin-polarized sample.  This is accomplished by precisely controlling the frequency and polarization of the Raman lasers.  Although the molecule has a body-fixed dipole moment $\mu$, the rovibrational ground state is an eigenstate of parity, and thus has no space-fixed dipole moment $d$ in the absence of an external field.  A DC electric field can be applied to mix opposite-parity rotational states and induce a dipole moment $d$ in the lab frame (the rotational states are labeled by $\ket{N,m_N}$, where $N$ is the principal rotational quantum number and $m_N$ is its projection on the quantization axis).  Experimentally, one can measure $d$ by performing Stark spectroscopy (Fig.~2c), which simply measures the energy shift of $\ket{g}$ in an electric field.  The electric field magnitude needed to saturate the dipole moment depends on the critical field $E_c$ ($E_c = B/\mu$, where $B$ is the rotational constant).  For example, to reach $d= 0.8 \mu$ for the $N=0$ state, a field of about 10 $E_c$ is required.

Another way to generate strong dipolar interactions in the lab frame is to directly couple two opposite-parity rotational states using AC electric fields \cite{Ospelkaus2010a}.  By coupling the $N=0$ and $N=1$ states with microwaves, one introduces dipolar interactions arising from an oscillating dipole moment of $\mu/\sqrt{3}$.  Figure~2d shows a simplified schematic of the rotational states of KRb.   Each rotational state $\ket{N,m_N}$ has 36 hyperfine states (since $I_\text{K}=4$ and $I_\text{Rb}=3/2$).  These states can be labeled by additional quantum numbers for the nuclear spin of K and Rb, $m_\text{K}$ and $m_\text{Rb}$, respectively.  A coupling between rotation and the nuclear electric quadrupole moment mixes these states, so $m_\text{K}$, $m_\text{Rb}$, and $m_\text{N}$ are not individually conserved; however, their sum $m_F=m_\text{K}+m_\text{Rb}+m_\text{N}$ is conserved \cite{Aldegunde2008}.  The strongest rotational transitions from the $\ket{0,0}$ state are those that preserve $m_\text{K}$ and $m_\text{Rb}$. To change the hyperfine state in the ground rovibrational state, one can take advantage of the aforementioned coupling in the $N=1$ manifold to perform a two-photon transition, where two hyperfine states in the $N=0$ manifold are coupled through a common state in the $N=1$ manifold \cite{Ospelkaus2010a, Will2016, Gregory2016}. We note that direct magnetic dipole transitions between hyperfine states in the ground state are very weak. Sample data for such two-photon transitions are shown in Fig.~2e.  These two-photon transitions have many applications: they can be used to transfer the molecules to the absolute lowest energy hyperfine state in cases where STIRAP does not initially populate the lowest energy state \cite{Ospelkaus2010a}, or they can be used to prepare long-lived coherent superpositions of two hyperfine states \cite{Park2016} (Fig.~2g).

When working with molecules confined in an optical trap, an important issue for preserving coherence of the $N=0$ and $N=1$ superposition is to match the AC polarizability of the two rotational states at the optical trapping wavelength \cite{Neyenhuis2012}.  The motivation for doing so is the same as operating an optical lattice clock at the ``magic wavelength" \cite{Ye2008}.  This can be achieved by some combination of electric and magnetic fields, as well as by the orientation of the light polarization with respect to the quantization axis \cite{Kotochigova2010}.  In Ref.~\cite{Neyenhuis2012}, the real part of the polarizability of the $N=0$ and $N=1$ states was measured as a function of the angle between the polarization vector of a vertically oriented lattice and a magnetic field (Fig.~2f), in the absence a DC electric field.  This led to the determination of the ``magic angle," which is especially pronounced for the $\ket{0,0} - \ket{1,0}$ transition.  At the ``magic angle," the coherence time of rotational superpositions is maximized. However, due to hyperpolarizability (the dependence of the polarizability on intensity), the coherence time was limited by the variation in light intensity across the molecular cloud.

\textbf{Controlling chemical reactions.} For many applications based on ultracold molecules, it is mandatory for the system to be stable and long-lived.  In spin-polarized atomic systems, the dominant loss mechanism is three-body recombination, where two atoms collide to form a weakly-bound molecule and the third atom simultaneously carries away the binding energy, leading to the loss of all three.  With polar molecules, two-body chemical reactions can lead to loss; for example, the reaction 2AB $\rightarrow$ A$_2$+B$_2$ can be exothermic.  A simple model that describes these chemical reactions is based on multichannel quantum-defect theory (MQDT) \cite{Idziaszek2010}, and assumes that once two molecules get to sufficiently short range, they react with high probability.  In the limit that this probability is unity, the chemical reactions are universal and can be described with knowledge of only the long-range part of the potential.  For spin-polarized fermions, the lowest energy collision channel is $p$-wave, giving rise to a centrifugal energy barrier that suppresses the rate at which molecules come to short range (Fig.~3a), while bosons (or distinguishable particles) can collide via $s$-wave interactions, leading to much higher reaction rates than for $p$-wave collisions.

Fermionic KRb molecules were found to undergo exothermic chemical reactions at ultralow temperatures \cite{Ospelkaus2010b, Ni2010}.  These chemical reactions can be described with simple quantum collision processes involving a few partial waves (Figs.~3b,c).  The anisotropy of the dipole-dipole interactions plays a crucial role.  In an applied electric field, molecules colliding in the ``head-to-tail" orientation have a lower $p$-wave barrier than molecules colliding in the repulsive ``side-by-side" orientation.  Thus, ``head-to-tail" collisions are the dominant loss channel, and can be suppressed by placing the molecules in a 1D lattice to create effective two-dimensional optical traps with a DC electric field applied perpendicularly to the plane \cite{Miranda2011}, or shut off completely by confining the molecules in individual zero-dimensional sites of a 3D lattice \cite{Chotia2012} (Fig.~3d).  In a 1D lattice, both the quantum statistics and lattice band distribution of the molecules play an important role (Fig.~3e).  Because of finite temperature and finite harmonic confinement, the experiments of Ref.~\cite{Miranda2011} were not performed in a fully two-dimensional trap, and thus the inelastic loss rate was only suppressed by about a factor of 60.  For very strong optical confinements and sufficiently large electric fields, repulsive dipolar interactions can suppress inelastic collisions, regardless of the quantum statistics \cite{Buchler2007, Micheli2007}.

In a 3D lattice, if $s$-wave inelastic collisions are allowed, the onsite loss rate $\Gamma$ for two molecules is very large.  $\Gamma$ can be significantly larger than the tunneling rate $J$ and be of the same order of magnitude as the energy gap between the two lowest lattice bands.  In this regime, molecular tunneling onto a lattice site that is already occupied by other molecules is suppressed by the quantum Zeno effect, and, counterintuitively, the effective loss rate of the whole system scales as $J^2/\Gamma$, decreasing as the onsite loss rate increases (see Fig.~3e, iii).  This continuous quantum Zeno suppression was observed first in homonuclear Rb$_2$ Feshbach molecules \cite{Syassen2008} and more recently with an incoherent mixture of two rotational states in KRb \cite{Yan2013, Zhu2014}.

For molecules that are chemically stable with respect to two-body loss, another proposed loss mechanism is three-body loss via ``sticky" collisions \cite{Croft2014, Mayle2013}, where two molecules first collide to form a reaction complex.  While they are close to each other, a third molecule approaches and forces the complex to a more deeply bound molecular state, leading to inelastic loss of all three molecules.  Preliminary experiments with RbCs \cite{Takekoshi2014} and NaRb \cite{Guo2016} suggest this may actually occur (Fig.~3f), although more work is needed to confirm this.

\textbf{Quantum magnetism with polar molecules.}  Turning off the chemical reactions in KRb in the 3D lattice made it possible to observe the long-range nature of the dipole-dipole interaction between molecules pinned in a deep optical lattice where tunneling, and thus both elastic and inelastic contact interactions, are absent.  Polar molecules can be used to study quantum magnetism where an interacting spin-$1/2$ Hamiltonian can be constructed based on the use of two opposite-parity rotational states~\cite{Barnett2006, Gorshkov2011b}.  Specifically, the experiments of Refs.~\cite{Yan2013, Hazzard2014} used the following mapping: $\ket{\da} \equiv \ket{0,0}$ and $\ket{\ua} \equiv \ket{1,-1}$ or $\ket{\ua} \equiv \ket{1,0}$.  The molecules undergo spin-exchange interactions, depicted schematically in Fig.~4a, and described by a long-range XY model:
\begin{equation}\label{eqn:goal}
H = \frac{1}{2} \sum_{i \neq j} V_{dd}(\mathbf{r_i}-\mathbf{r_j})  \frac{J_\perp}{2} \left(\hat{S}_i^{+} \hat{S}_j^{-} + \hat{S}_i^{-} \hat{S}_j^{+}  \right),
\end{equation}
where $V_{dd}(\mathbf{r_i}-\mathbf{r_j})= \frac{1-3 \cos^2 \theta_{ij}}{|\mathbf{r_i}-\mathbf{r_j}|^3}$, $\mathbf{r_i}$ is the position of molecule $i$ in units of the lattice spacing, $\theta_{ij}$ is the angle between $\mathbf{r_i}-\mathbf{r_j}$ and the quantization axis, and $\hat{S^+}$ and $\hat{S^-}$ are spin-$1/2$ raising and lowering operators.  The coupling constant $J_\perp \propto \frac{d_{\da \ua}^2}{a_\text{lat}^3}$, where $d_{\da \ua}$ is the transition dipole moment between the two rotational states (equal to $\mu/\sqrt{3}$ at zero electric field) and $a_\text{lat}$ is the lattice constant.  In an applied DC field, there is also an Ising interaction $\sim \sum_{i,j} J_z \hat{S_i^z} \hat{S_j^z}$.  Additionally, there are other terms omitted from Eq.~\ref{eqn:goal} that can convert between rotational and orbital angular momentum, which can give rise to a novel form of spin-orbit coupling \cite{Syzranov2014}.

In Refs.~\cite{Yan2013, Hazzard2014}, the spin dynamics were probed with Ramsey spectroscopy to measure the time evolution of the spin coherence (Fig.~4b).  A spin echo was required to mitigate the effects of rapid single-particle dephasing resulting from residual differential AC Stark shifts \cite{Neyenhuis2012}.  The contrast decay was observed to have a dependence on both the molecular density and on the choice of excited state ($J_{\perp}$ is a factor of two larger for $\ket{\ua} = \ket{1,0}$ compared to $\ket{\ua} = \ket{1,-1}$).  Theoretical modeling based on a cluster expansion showed that interactions beyond nearest and next-nearest neighbor are necessary to explain the experimental data, and estimated the lattice fillings to be 5-10\% \cite{Hazzard2014}.  At higher lattice fillings and with improved imaging resolution, such a system shows great potential as a quantum simulator that could address questions in quantum many-body dynamics that are computationally intractable.  We note that similar experiments have been performed in magnetic Cr atoms \cite{dePaz2013}.

\textbf{Creating a dense molecular gas in a 3D lattice}.  The difficulty of cooling chemically reactive molecules in bulk, combined with the stability afforded by the 3D lattice \cite{Chotia2012}, suggests an alternative way to create a low entropy gas of polar molecules: to load atom gases into the lattice to optimize the distribution of each atomic species and then make molecules at individual lattice sites~\cite{Damski2003, Freericks2010}.  In the case of a Bose-Fermi mixture, the ideal distributions are a Mott insulator (MI) for the bosons \cite{Greiner2002} and a spin-polarized band insulator for the fermions \cite{Schneider2008}, with precisely one atom of each species per site in the center of the lattice \cite{Moses2015}.  This parallels earlier work with homonuclear Rb$_2$ and Cs$_2$ molecules, where molecules were created from a MI in a 3D lattice \cite{Danzl2010, Volz2006}, optimized to have two atoms per site in the center.  In these experiments, the density overlap was very simple since only a single atomic species was involved.  However, in the Bose-Fermi case, the situation is very different, as the two species have vastly different requirements to have one particle per site: high trap frequencies (large compression) and large atom numbers for the fermions, but low trap frequencies and small atom numbers for the bosons.  In the Bose-Bose case, the challenge is to spatially overlap two single-atom Mott insulators.

Refs.~\cite{Moses2015, Covey2016} systematically studied this quantum synthesis approach for the $^{40}$K$^{87}$Rb Bose-Fermi mixture (Fig.~4c,d).  The first step involved determining the filling of the atomic gases by themselves, and indeed it was found that a small Rb quantum gas (with less than 5000 atoms) and a large K degenerate gas (with more than $10^5$ atoms) were required to achieve fillings approaching 1 per site at the center. The next step was to ensure an optimal spatial overlap between these two low-entropy atomic distributions and to guard against any negative impact of the presence of one species on the other.  In order to preserve the filling of the Rb MI in the presence of such a large K cloud, it was imperative to turn off the interspecies interactions by loading the lattice at $a_\text{K-Rb}=0$.  In the limit of small Rb atom numbers, magnetoassociation converted more than 50\% of the Rb to Feshbach molecules.  Converting Feshbach molecules into ground-state molecules, fillings $> \, $25\% in a 3D lattice were achieved, significantly higher than previous results on ground-state molecules \cite{Yan2013, Hazzard2014}, and close to the percolation threshold, where every molecule is connected to every other molecule \cite{Moses2015}.  Future experimental improvements based on the findings of Ref.~\cite{Covey2016} should be able to significantly increase the filling.  Similar recent work created a sample of bosonic RbCs Feshbach molecules, at a lattice filling $> \,$30\% \cite{Reichsoellner2016}.  There, the Cs was first localized in the MI phase, the Rb was next translated to overlap with the Cs cloud, the lattice depth was then increased further to localize Rb, and finally RbCs Feshbach molecules were created via magnetoassociation, setting the stage to produce ground-state molecules \cite{Reichsoellner2016}.  
  
\textbf{Outlook.}  The recent proliferation in the number of groups capable of producing ultracold polar molecules demonstrates the vibrancy of the field and suggests rapid progress in the near future.  The scope of the field may expand in a number of different directions.  One direction is to improve the optical and electric field control over the molecules.  As an example, a new generation of the KRb experiment is currently being built, which features transparent electrodes in the vacuum chamber, and a high-NA objective for $\sim 1 \, \mu$m resolution (Fig.~4e).  These features should enable precise tuning of the coupling constants of the general XXZ Hamiltonian \cite{Hazzard2013} and allow direct experimental observations of the propagation of correlations and entanglement (or lack thereof, as in many-body localized phases \cite{Yao2014}).  Similar experiments are being constructed in other groups around the world.  Ultimately, one would like to build a quantum gas microscope for molecules (Fig.~4e), similar to those implemented for neutral atoms \cite{Haller2015, Cheuk2015, Parsons2015, Omran2015}.  Another direction is to produce new species of molecules, such as KCs \cite{Grobner2016}, NaCs \cite{Hutzler2016} and open shell molecules such as RbSr \cite{Zuchowski2010, Pasquiou2013}, LiYb \cite{Dowd2015}, and CsYb \cite{Kemp2016}.

Recent work has been especially encouraging.  For example, the NaRb experiment has induced a dipole moment of 1.06 D \cite{Guo2016}, corresponding to a dipole length of $\sim 1.9 \, \mu$m, which is more than 3 lattice sites in a lattice with 532 nm spacing.  Such large interactions with bosons in lattices should lead to novel studies of the extended Bose-Hubbard model, possibly paralleling the work in Ref.~\cite{Baier2016} with highly magnetic atoms.  Dipolar interactions modify the traditional Hubbard model and can lead to novel phases including supersolid and stripe phases \cite{Yi2007, Pollet2010, Sansone2010}.  In bulk gases of bosonic magnetic atoms, the competition between contact and dipolar interactions has led to the observation of droplets, which are stabilized by quantum fluctuations \cite{FerrierBarbut2016, Chomaz2016}.  For fermionic molecules, the dipolar interactions can contribute an energy comparable to the Fermi energy \cite{Park2015}, and can give rise to topological superfluid phases \cite{Baranov2002, Cooper2009} and superfluid pairing between layers of an optical lattice \cite{Pikovski2010}.  The field of ultracold polar molecules shows no signs of slowing down, and there should be many fruitful experiments in the next few years.

\noindent \textbf{Acknowledgements.}  This article is dedicated to the memory of Deborah Jin, who passed away on 15 September 2016 after a courageous battle with cancer. Debbie was a beloved friend, colleague, and teacher. She demonstrated an unparalleled combination of scientific vision, creativity, and detail-oriented experimental excellence. Among her many outstanding accomplishments, Debbie was a guiding force on the JILA KRb polar molecule collaboration for the past dozen years. Her vision is clearly manifest in the legacy of our work. Her ideas and sense of direction for our experiment will continue to influence our work for many years to come. Even when we progress sufficiently far on the experiment beyond anything we could have imagined during her time, we will continue to feel inspired by her creativity and enthusiasm. We, and the entire physics community, will deeply miss her. 

\noindent \textbf{Author contributions.}  All authors contributed to the writing of the manuscript.

\noindent \textbf{Author information.}  The authors declare no competing financial interests.  Correspondence and request for materials should be sent to J. Y. (ye@jila.colorado.edu).

\begin{figure}
\begin{center}
\includegraphics[width=16cm]{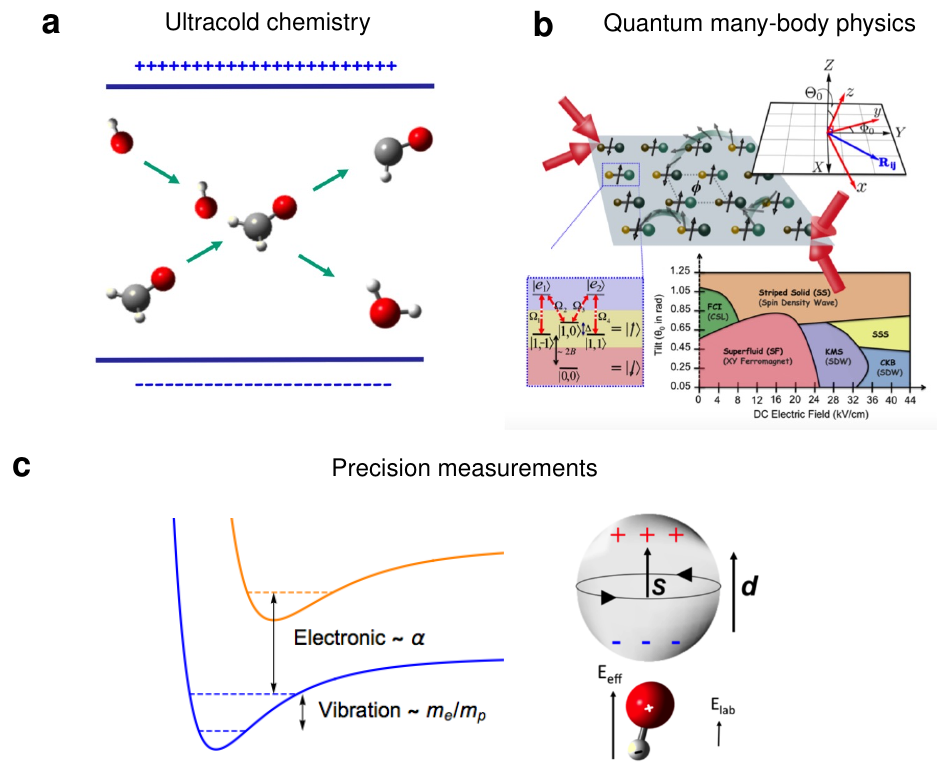}
\caption{ \textbf{A survey of the applications of ultracold molecules.  a},  Working with ultracold molecules gives experimenters the ability to precisely control the initial states of a chemical reaction, determine the rovibrational states of the products, and use external electric and magnetic fields to control reaction rates and pathways.  \textbf{b}, Polar molecules are ideal candidates for realizing novel quantum many-body systems and for encoding quantum information (in rotational or hyperfine states, for example) \cite{Demille2002}.  The figure illustrates a proposal to realize a fractional Chern insulator with a 2D array of polar molecules.  Reprinted figure with permission from Ref.~\cite{Yao2013}.  Copyright (2013) by the American Physical Society.  \textbf{c}, Because of their many degrees of freedom, polar molecules can be used for a variety of precision measurement applications \cite{Hansch2006}, including tests for parity and CP violation \cite{Kozlov1995} and time variation of fundamental constants such as the fine structure constant (for which electronic transitions are more sensitive) \cite{Hudson2006} and the electron-to-proton mass ratio \cite{Zelevinsky2008} (for which vibrational transitions are more sensitive).  Polar molecules are especially favorable for electron electric dipole moment searches because a small electric field in the lab produces a huge effective electric field in the molecule \cite{Hudson2011, Loh2013, Acme2014}.}
\label{fig:applicationsfig}
\end{center}
\end{figure}

\begin{figure}
\begin{center}
\vspace{-2cm}
\includegraphics[width=16cm]{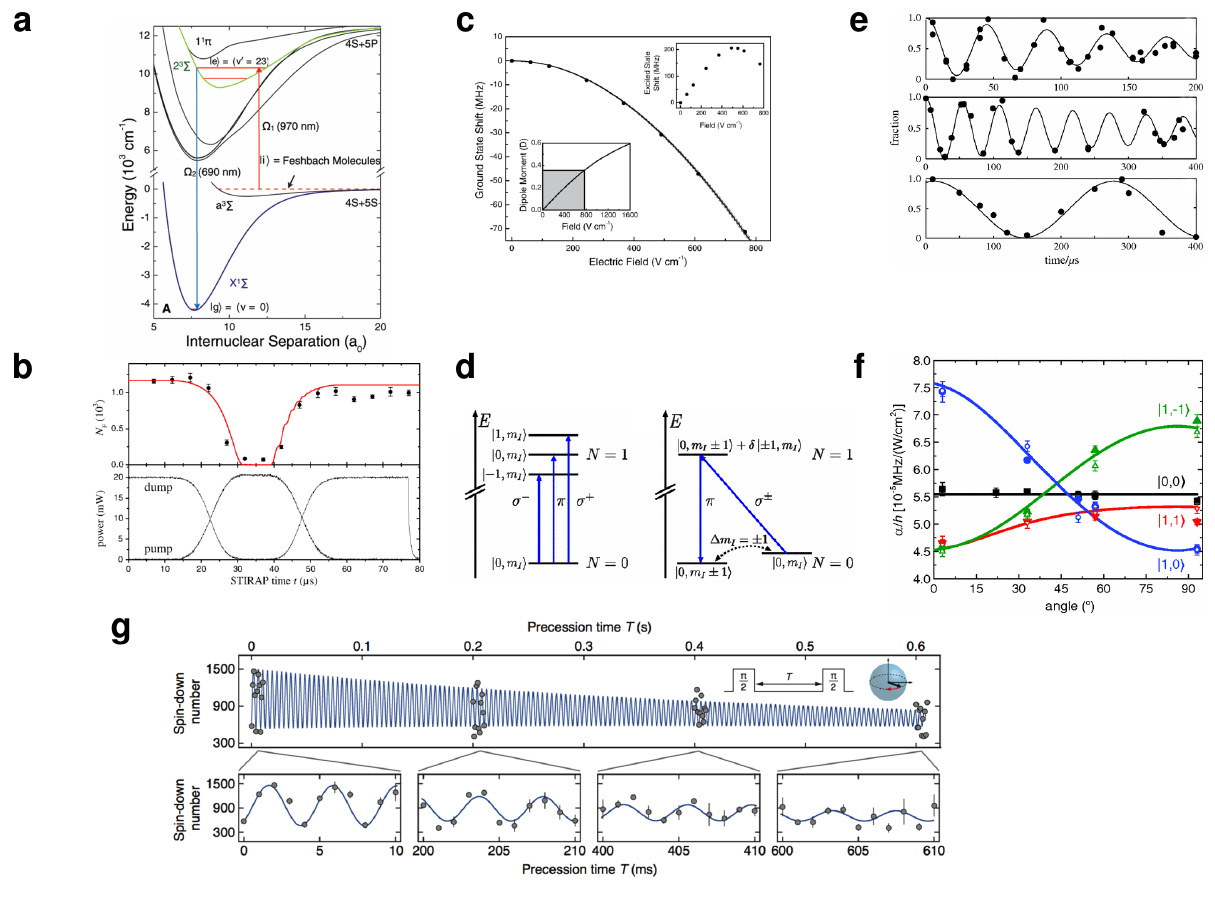}
\vspace{-0.5cm}
\caption{ \textbf{Creating and manipulating ultracold bialkali molecules.  a}, The energy levels of KRb molecules, from Ref.~\cite{Ni2008}.  Reprinted with permission from AAAS.  STIRAP couples the Feshbach state with the ground state via an electronic excited state (see text).  \textbf{b}, The upper panel plots the number of Feshbach molecules vs.~time during the STIRAP sequence, while the bottom panel plots the laser intensities.  The ground-state molecules are dark to the Feshbach molecule detection.  Reprinted figure with permission from Ref.~\cite{Takekoshi2014}.  Copyright (2014) by the American Physical Society.  \textbf{c}, Stark shift of RbCs ground-state molecules, from which a dipole moment of 1.23 Debye is inferred.  The inset shows the Stark shift of the excited state.  Reprinted figure with permission from Ref.~\cite{Molony2014}.  Copyright (2014) by the American Physical Society.  \textbf{d}, The lowest energy rotational states of KRb.  To couple hyperfine states in the $N=0$ manifold, a two-photon Raman transition via the $N=1$ state is used.  Reprinted figure with permission from Ref.~\cite{Ospelkaus2010a}.  Copyright (2010) by the American Physical Society.  \textbf{e}, Rabi oscillations between rotational states in KRb.  In the top panel, the hyperfine state is unchanged, while in the middle and bottom panels, $m_\text{Rb}$ and $m_\text{K}$ are changed, respectively.  Reprinted figure with permission from Ref.~\cite{Ospelkaus2010a}.  Copyright (2010) by the American Physical Society.    \textbf{f}, The real part of the polarizability of KRb molecules at 1064 nm vs.~the angle between a linearly polarized lattice beam and the quantization axis.  When the polarizabilities of two states are equal, a ``magic" condition leads to an enhanced coherence time for rotational superpositions.  Reprinted figure with permission from Ref.~\cite{Neyenhuis2012}.  Copyright (2012) by the American Physical Society.    \textbf{g}, A Ramsey experiment between two hyperfine states of the $N=0$ manifold of NaK reveals a coherence time $\sim 1$ second.  Figure reproduced with permission from Ref.~\cite{Park2016}.}
\label{fig:fig2}
\end{center}
\end{figure}

\begin{figure}
\begin{center}
\vspace{-0.4cm}
\includegraphics[width=16cm]{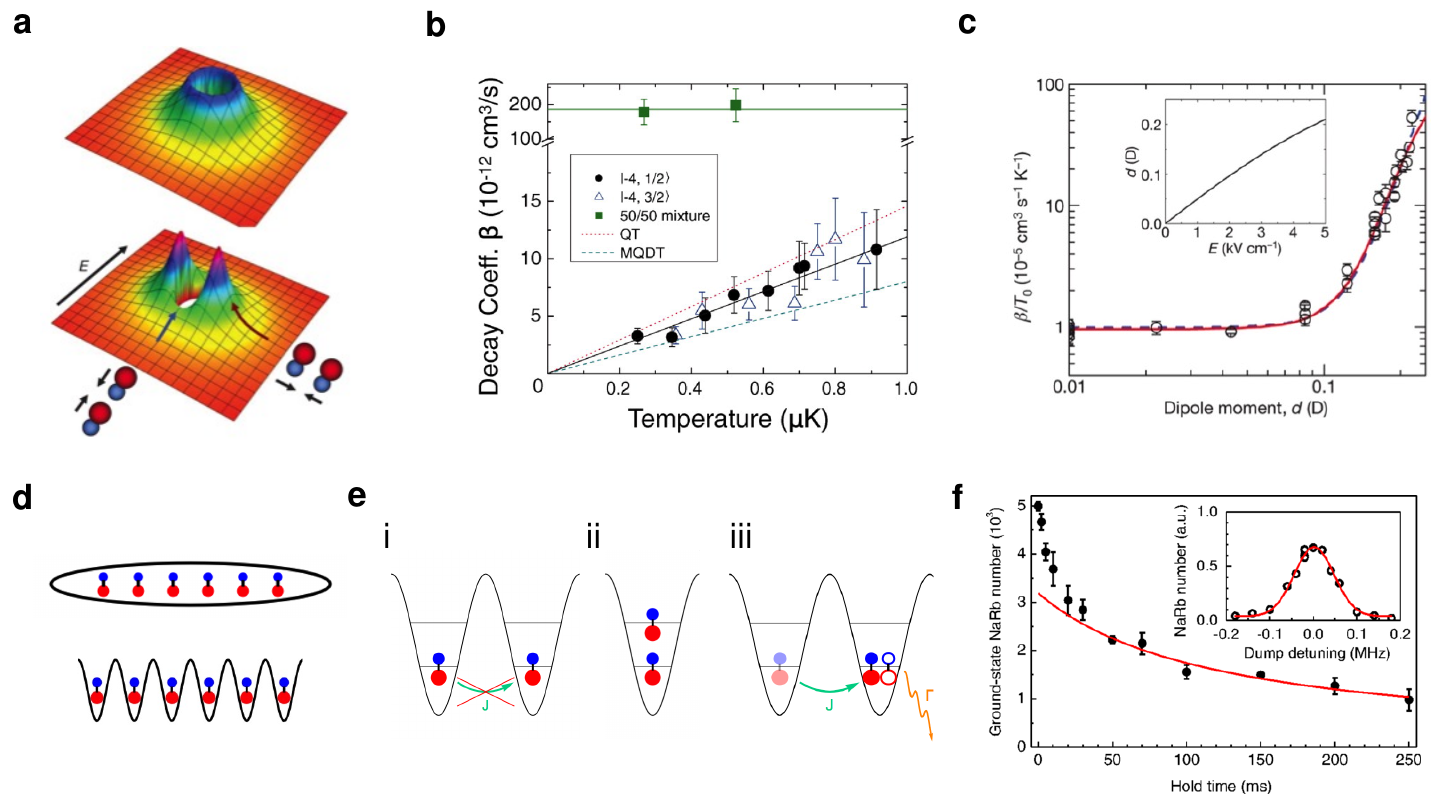}
\vspace{-0.4cm}
\caption{ \textbf{Controlling chemical reactions with optical lattices.  a}, 
At zero electric field, the $p$-wave centrifugal barrier for spin-polarized fermionic KRb molecules is isotropic (top), while in an applied DC field, the barrier is reduced for attractive ``head-to-tail" collisions and increased for repulsive ``side-by-side" collisions (bottom). Reprinted by permission from Macmillan Publishers Ltd: Nature \cite{Ni2010}, copyright (2010).  \textbf{b}, The temperature dependence of the chemical reaction rate at zero electric field, from Ref.~\cite{Ospelkaus2010b}.  Reprinted with permission from AAAS.  For spin-polarized samples, the two-body decay rate $\beta \propto T$, as expected from the Wigner threshold laws.  For a spin-mixture, the decay rate is much higher and independent of temperature.  \textbf{c}, The scaled chemical reaction rate for a spin-polarized sample vs.~induced dipole moment $d$ scales as $d^6$.  The inset shows $d$ vs.~applied electric field.  Reprinted by permission from Macmillan Publishers Ltd: Nature \cite{Ni2010}, copyright (2010).  \textbf{d}, By confining chemically reactive molecules in optical lattices, the reactions can be suppressed in a 1D lattice (upper picture) \cite{Miranda2011} or effectively shut off in a 3D lattice (lower picture) \cite{Chotia2012}.  \textbf{e}, The chemical reactivity in the lattice depends highly on the molecules' quantum statistics and band distribution.  (i) Pauli blocking prevents indistinguishable molecules in the lowest band from occupying the same lattice site, suppressing chemical reactions.  (ii) Indistinguishable molecules in different bands can chemically react in the ``head-to-tail" orientation \cite{Miranda2011}.  Molecules could be in higher bands because of heating or high initial temperature.  (iii) Distinguishable molecules can occupy the same lattice site and chemically react with no centrifugal energy barrier.  However, if the chemical reaction rate $\Gamma$ is much larger than the tunneling rate $J$, the quantum Zeno effect suppresses the chemical reactions \cite{Zhu2014}.   \textbf{f}, Even for molecules that are chemically stable to two-body chemical reactions, such as NaRb, there is another predicted loss mechanism due to ``sticky" collisions (see text).  Here, the loss of spin-polarized NaRb molecules in the absolute ground-state is plotted, along with a fit assuming two-body loss.  Reprinted figure with permission from Ref.~\cite{Guo2016}.  Copyright (2016) by the American Physical Society.}
\label{fig:chemistryfig}
\end{center}
\end{figure}

\begin{figure}
\begin{center}
\includegraphics[width=16cm]{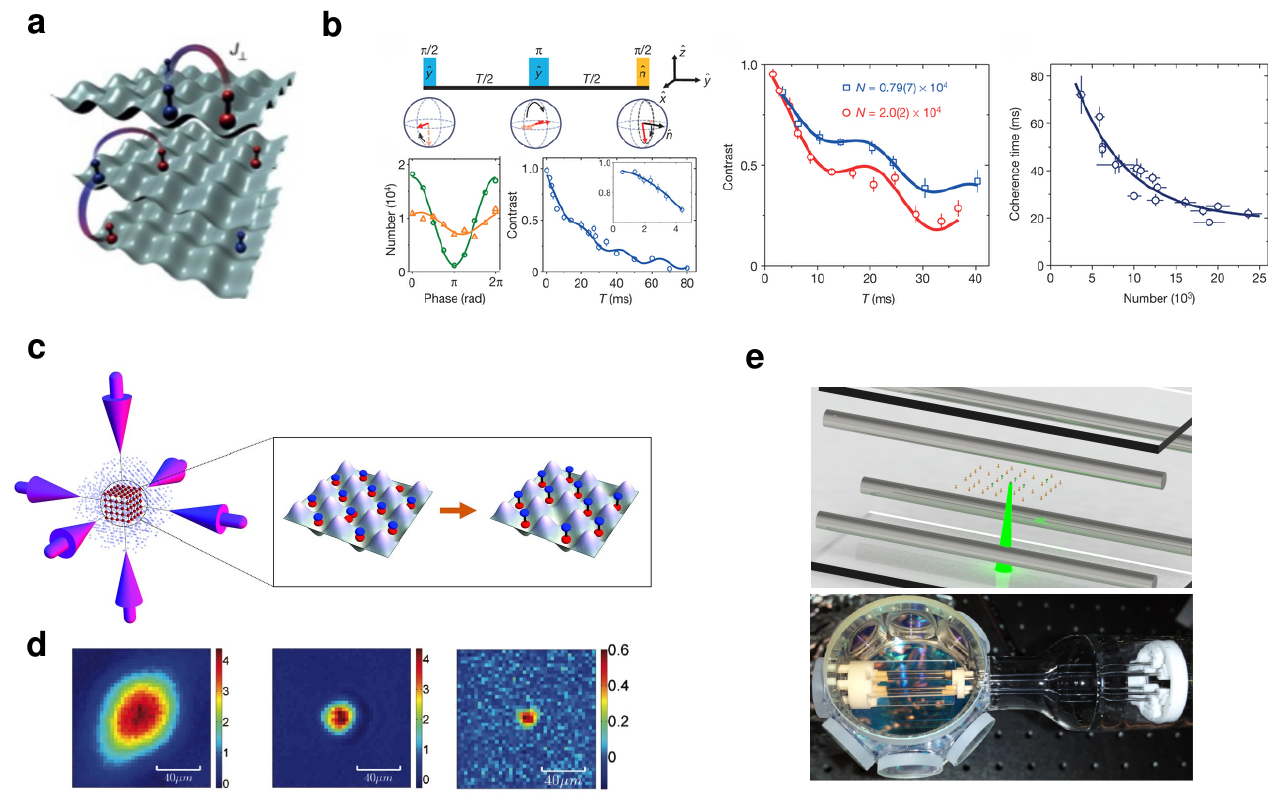}
\caption{ \textbf{Recent experiments with polar molecules in optical lattices.  a}, Schematic of spin exchange, from Ref.~\cite{Yan2013}.  \textbf{b}, Experimentally, a spin-echo Ramsey experiment was used to observe spin exchange.  The contrast decays in a density-dependent way (the coherence time $\tau \sim 1/N$) and oscillates roughly independently of density.  These experiments were performed at very low filling fractions, or densities \cite{Yan2013}.  \textbf{c}, The scheme employed in Ref.~\cite{Moses2015} to increase the molecule filling fraction was to first optimize the density and spatial overlap of the K (blue) and Rb (red) quantum gases.  Lattice sites with one K and one Rb are efficiently converted to molecules \cite{Chotia2012, Covey2016}.  \textbf{d}, Typical \textit{in situ} absorption images of K (left), Rb (middle), and ground-state molecules (right), from Ref.~\cite{Moses2015}.  The Rb cloud is significantly smaller than the K cloud in order to have a density of one atom per site in the center of the lattice.  The optimized ground-state molecule filling is more than 25\%.  \textbf{e}, A quantum gas microscope for molecules would allow for single-site spin-sensitive detection and manipulation (top).  The science cell of the second generation KRb experiment at JILA (bottom) features improved imaging resolution and better electric field control via indium-tin-oxide (ITO) coated plates and tungsten rods in vacuum, and is compatible with a high-NA objective for quantum gas microscopy.}
\label{fig:recentexps}
\end{center}
\end{figure}

\bibliographystyle{unsrt}
\bibliography{npreviewlib}

\end{document}